\documentclass{ws-procs9x6-cpt16}
\begin{document}

\newcommand{\refeq}[1]{(\ref{#1})}
\def\etal {{\it et al.}}

\def\ke{(\overline{k}_{\mathrm{eff}})}
\def\ko{(k_{1}^{(6)})}
\def\kt{(k_{2}^{(6)})}

\title{Search for Lorentz Violation using Short-Range Tests of Gravity}

\author{J.\ Long}

\address{Physics Department, Indiana University,
Bloomington, IN 47405, USA}

\begin{abstract}
Experimental tests of the newtonian inverse square law at short
range, one at Indiana University and the other at the Huazhong
University of Science and Technology, have been used to set limits on
Lorentz violation in the pure gravity sector of the nonminimal
Standard-Model Extension.  In the nonrelativistic limit, the
constraints derived for the 14 independent SME coefficients for 
Lorentz violation acting simultaneously are of order $\ke\sim
10^{-9}$~m$^{2}$.
\end{abstract}

\bodymatter

\section{Introduction}
Local Lorentz invariance is at the foundation of both
the Standard Model and General Relativity, but is not as well tested
for the latter theory. Violation of Lorentz symmetry would break the
isotropy of spacetime, permitting the vacuum to fill with
``background'' fields with a preferred direction.  Interaction of the
masses in a terrestrial gravity experiment with these fields could
result in sidereal modulations of the force between the masses,
providing a test of Lorentz invariance in gravity.  

A quantitative description of Lorentz violation consistent with local
field theory is given by the Standard-Model Extension (SME), which has
been expanded to include gravitational effects by introducing a Lagrange
density containing the usual Einstein-Hilbert term, plus an 
infinite series of operators of increasing mass dimension $d$
representing corrections to known physics at attainable
scales.\cite{ak04}  To date, the minimal ($d=4$) and
nonminimal cases have been investigated theoretically up to
$d=6$.\cite{Bailey}  In the nonrelativistic limit, the
expression for the $j$th component of the force 
between two point masses $m_{1}$ and $m_{2}$ in the nonminimal SME is
given by:  
\begin{equation}
F^{j} = -Gm_{1}m_{2}\left(\frac{\hat{R}^{j}}{R^{2}}-\frac{\overline{k}_{j}(\hat{R},T)}{R^{4}}\right),
\label{eq:LVgrav}
\end{equation}
where the first term is newtonian gravity
and the second is an SME correction.
Here, $\vec{R}$ is the vector separating
$m_{1}$ and $m_{2}$, and $\hat{R}^j$ is the projection of the unit vector
along $\vec{R}$ in the $j$th direction. 
The SME correction term is
\begin{eqnarray}
\overline{k}_{j}(\hat{R},T)&=&\frac{105}{2}\ke_{klmn}\hat{R}^{j}\hat{R}^{k}\hat{R}^{l}\hat{R}^{m}\hat{R}^{n}-45\ke_{klmn}\hat{R}^{j}\hat{R}^{k}\hat{R}^{l}
\nonumber \\&& +\frac{9}{2}\ke_{klkl}\hat{R}^{j}-30\ke_{jklm}\hat{R}^{k}\hat{R}^{l}\hat{R}^{m}+18\ke_{jkll}\hat{R}^{k},
\end{eqnarray}
where $\ke_{jklm}$ contains 14 independent coefficients
for Lorentz violation with units m$^{2}$ in the standard laboratory frame.
Motivated by Ref.\ \refcite{Bailey}, this report presents a test of
Eq.~\refeq{eq:LVgrav} in laboratory gravity experiments.

\section{The Indiana short-range experiment}
The Indiana experiment is optimized for sensitivity to macroscopic
forces beyond gravity at short range, which in turn could
arise from exotic elementary particles or even extra spacetime
dimensions.  It is described in detail elsewhere;\cite{Long03,Yan14}
here we concentrate on the essential features.    

The experiment is illustrated in Fig.~1 of Ref.\ \refcite{Long15}. The
test masses consist of 250~$\mu$m thick planar tungsten 
oscillators, separated by a gap of 100~$\mu$m, with a stiff conducting
shield in between them to suppress 
electrostatic and acoustic backgrounds.  Planar geometry
concentrates as much mass as possible at
the scale of interest, and is nominally null with
respect to $1/r^{2}$ forces.  This is effective in suppressing the
newtonian background relative to exotic short-range effects, and would
be expected to be ideal for testing Eq.~\refeq{eq:LVgrav}, in which the
SME correction term varies as $1/r^{4}$. The force-sensitive
``detector'' mass is driven by the force-generating ``source'' mass
at a resonance near 1~kHz, placing a heavy burden on
vibration isolation.  The 1 kHz operation is chosen since at this
frequency it is possible to construct a simple vibration
isolation system. This design has proven effective for suppressing all
background forces to the extent that the only
effect observed is thermal noise due to dissipation in the detector
mass.\cite{Yan14}  After a run in 2002, the experiment set the
strongest limits on forces beyond gravity between 10 and
100~$\mu$m.\cite{Long03}  The experiment has since been optimized to
explore gaps below 50~$\mu$m, and new force data were acquired in
2012.

Analysis of the 2002 and 2012 data for evidence of Lorentz violation
requires a theoretical expression for the Lorentz violating force for
the particular geometry.  Equation \refeq{eq:LVgrav} is evaluated
by Monte Carlo integration with the geometrical parameters listed in
Refs.\ \refcite{Long03} and \refcite{Long15}. The result can be
expressed as a Fourier series of the time dependence,
\begin{equation}
F=C_{0}+\sum_{m=1}^{4}S_{m}\sin(m\omega_{\oplus}T)+C_{m}\cos(m\omega_{\oplus}T), 
\label{eq:LVF}
\end{equation}
where $C_{m}, S_{m}$ are functions of the SME coefficients and test
mass geometry (and the laboratory colatitude angle).  In
Eq.~\refeq{eq:LVF}, the SME coefficients are expressed in the
Sun-centered celestial equatorial frame, which are related to the 
laboratory-frame coefficients by $\ke_{jklm}=M^{jJ}M^{kK}M^{lL}M^{mM}\ke_{JKLM}$, where
the matrix $M$ is given by Eq.~(10) of Ref.\ \refcite{Bailey}.  The term
$\omega_{\oplus}$ is the Earth's sidereal rotation frequency,
and the time $T$ is measured in the Sun-centered frame. 
The result for the constant term is:
\begin{eqnarray}
C_{0}&=&[-(1.8\pm 2.3)\ke_{XXXX}-(1.8\pm 2.3)\ke_{YYYY} \nonumber \\
&&-(3.6\pm 4.7)\ke_{XXYY}+(13.5\pm 7.5)\ke_{XXZZ} \nonumber \\
&&-(13.5\pm 7.5)\ke_{YYZZ}]~\mathrm{nN/m^{2}}.
\label{eq:c0}
\end{eqnarray}
Each term is quite sensitive to uncertainties in the test mass
geometry (which determine the errors); in fact the means are smaller
than might be expected for a simple $1/r^{4}$ force for planar geometry. 

Some insight can be gained from examination of the force in
Eq.~\refeq{eq:LVgrav} for the case of a point mass $m$ suspended a
distance $d$ above the
center of a circular plate of radius $\rho$, which can be solved
analytically.  The result can be expressed as $1/d^{2}$ times a linear
combination of oscillatory angular functions
$\Gamma^{jklm}(\theta,\phi)$, each function weighted by an SME laboratory-frame
coefficient.  Here, $\theta$ and $\phi$ are the polar and azimuthal
angles of the vector between $m$ and a mass element $dm$ in the
plate.  In particular, nine of the $\Gamma$ vanish upon integration of
$\phi$ over $2\pi$ radians.  The remaining terms vanish upon integration of
$\theta$ from 0 to $\pi/2$ (the case of an infinite plate), and are
strongly suppressed for $\rho /d > 8$, as the oscillatory structure of
the $\Gamma$ averages out.  Thus, the force in Eq.~\refeq{eq:LVgrav} is
suppressed in geometries with high symmetry and which subtend large
solid angles; both are characteristics of the geometry of the IU experiment. 

\section{Limits on Lorentz violation coefficients}
Analysis of the 2002 and 2012 data sets for signals of Lorentz
violation has been completed, following Ref.\ \refcite{Long15}. Time
stamps in the data are extracted and offset 
relative to the effective $T_{0}$ in the Sun-centered frame (taken to
be the 2000 vernal equinox).  Discrete Fourier transforms of
the data at each frequency component of the signal ($0, \omega_{\oplus},
2\omega_{\oplus}, 3\omega_{\oplus}, 4\omega_{\oplus}$) are computed,
with errors, and corrected for discontinuous time data.  Results,
shown in Table~I of Ref.\ \refcite{Long15}, are consistent with zero
with errors of order $\sim$10~fN.  Gaussian probability
distributions at each signal frequency component
are constructed, using the difference between the Fourier transforms
and the predicted signals (e.g., Eq.~\refeq{eq:c0}) as the means.  A
global probability distribution is constructed from the product of the
18 component distributions.  Means and errors of particular
$\ke_{JKLM}$ (for example, $\ke_{XXXX}$) are then computed by
integration of the distribution over all $\ke$ except
$\ke_{XXXX}$. Results are of order $\ke \le 10^{-5}$~m$^{2}$. 

Results improve significantly on inclusion of data from the
short-range experiment at the Huazhong University of Science and
Technology, a torsion balance with planar test masses separated by
$\sim 300$~$\mu$m.\cite{Tan16}  Terms in the Lorentz-violating
torque for this experiment (the analog of Eq.~\refeq{eq:c0}) are of order
10~nNm/m$^{2}$, while measured torque Fourier components have
errors $\sim 10$~aNm.  Improvement in sensitivity by a factor of
$\sim$10$^{3}$ would be expected; the resulting constraints on
the $\ke$ are typically $10^{-9}$~m$^{2}$.\cite{Shao16}  The $\ke$ are
derived from the 336 coefficients $\ko$ and $\kt$ in the fully
relativistic SME; the simultaneous constraints on $\ke$ translate into
comparable constraints on 131 fundamental coefficients taken one at a time. 

Further study of Eq.~\refeq{eq:LVgrav} reveals that (i) the
total force between point masses contains six components (each
weighted by a separate $\ke$) directed along the vector between the
masses with the remaining components directed {\it orthogonally}, (ii) variation
of the force in a terrestrial experiment is maximized when
the sensitive axis is orthogonal to the Earth's rotation axis,
and (iii) the point-plate force tends to a maximum when the separation of
the masses is on the order of the plate radius.  Future
experiments taking advantage of these features are expected to
have greater sensitivity.

\section*{Acknowledgments}
The author thanks R.\ Xu, A.\ Kosteleck\'y, and H.-O.\ Meyer for essential
discussions of Eq.~\refeq{eq:LVgrav}, and E.\ Weisman for help with
numerical calculations.
This work was supported in part by the Indiana University
Center for Spacetime Symmetries.

\end{document}